\documentclass[epj]{svjour}
\usepackage{graphics}
\usepackage{times}
\usepackage{helvet}
\begin{document}
\title{Azimuthal Anisotropy of Photon and Charged Particle Emission
in $^{208}$Pb + $^{208}$Pb Collisions at 158 $\cdot A$ GeV/$c$}
\subtitle{WA98 Collaboration}
\author{
M.M.~Aggarwal\inst{4}
\and Z.~Ahammed\inst{2}
\and A.L.S.~Angelis$^*$\inst{7} 
\and V.~Antonenko\inst{13}
\and V.~Arefiev\inst{6}
\and V.~Astakhov\inst{6}
\and V.~Avdeitchikov\inst{6}
\and T.C.~Awes\inst{16}
\and P.V.K.S.~Baba\inst{10}
% ...
}
\institute{
Institute of Physics, 751-005  Bhubaneswar, India
\and Variable Energy Cyclotron Centre,  Calcutta 700 064, India
\and MIT Cambridge, MA 02139, USA 
\and University of Panjab, Chandigarh 160014, India
\and Gesellschaft f{\"u}r Schwerionenforschung (GSI), D-64220 Darmstadt, Germany 
\and Joint Institute for Nuclear Research, RU-141980 Dubna, Russia
\and University of Geneva, CH-1211 Geneva 4,Switzerland
\and KVI, University of Groningen, NL-9747 AA Groningen, The Netherlands 
\and University of Rajasthan, Jaipur 302004, Rajasthan, India
\and University of Jammu, Jammu 180001, India
\and University of Tennessee, Knoxville, Tennessee 37966, USA
\and Lund University, SE-221 00 Lund, Sweden 
\and RRC ``Kurchatov Institute'', RU-123182 Moscow, Russia
\and University of M{\"u}nster, D-48149 M{\"u}nster, Germany
\and SUBATECH, Ecole des Mines, Nantes, France
\and Oak Ridge National Laboratory, Oak Ridge, Tennessee 37831-6372, USA
% ...
}

\abstract{
The azimuthal distributions of 
photons and charged particles with respect to the event plane
are investigated as a function of  centrality
in $^{208}$Pb + $^{208}$Pb 
collisions at 158 $\cdot A$ GeV/$c$ in the WA98 experiment
at the CERN SPS. 
The anisotropy of the azimuthal distributions is characterized using
a Fourier analysis.  For both the photon and charged particle
distributions the first two Fourier
coefficients are observed to decrease with increasing centrality. 
The observed anisotropies of the photon distributions compare well with 
the expectations from the charged particle measurements for all centralities.
}

\maketitle
\section{Introduction}

Non-isotropic 
emission of particles with respect to the reaction plane,
as first observed at the Bevalac ~\cite{bevalac},
provides evidence for collective flow in high energy heavy ion 
collisions. 
Flow, or anisotropic particle emission, has been observed
for a large variety of interacting
systems from incident energies of a few A GeV/$c$ at the Bevalac (SIS) 
and AGS to much greater energies at the SPS and RHIC
~\cite{ritter,eos,fopi,kaos,e877,e802,wa93,wa98,na49,star,phenix,phobos}.

Anisotropic flow manifests itself as asymmetries in the 
azimuthal distribution of particles and can be reproduced  in
theoretical models with different underlying assumptions. 
One scenario is incorporated in transport models where the particles
have a mean free path comparable to the system size~\cite{trans1,trans2,heisel}.
The models can describe the observed flow up to AGS energies. 
The other scenario applies when the mean free path is 
much smaller than the system size which allows 
the description of the equilibrated system in terms of 
macroscopic quantities~\cite{ollie1,kolb}.
Hydrodynamic models are able to describe the qualitative features 
of the observed flow~\cite{hydro} for p$_T$ below about 3 GeV/c.

The initial asymmetry 
in the overlap zone of the colliding nuclei translates into unequal
pressure gradients in different directions that leads 
to an elliptic final state momentum distribution of the 
particles~\cite{ollie1}, causing an elliptic pattern of flow. 
The elliptic flow is therefore expected to be sensitive 
to the system evolution at the time of maximum 
compression ~\cite{sorge1} and is 
shown to be sensitive to 
the equation of state of the compressed nuclear
matter. The variation of asymmetry with centrality
enables to relate the observed flow to the geometry of the overlap region 
~\cite{sorge2,heisel}. One would then expect 
a scaling of the data from AGS to SPS and RHIC provided the physics 
of elliptic flow remains the same ~\cite{volos99}. In the case that
there is a phase transition from hadronic matter to a quark gluon 
plasma, it is expected that the reflection of this transition in the equation
of state of the dense nuclear matter would result in changes in the 
pressure gradients which would then be reflected in changes in the particle
flow pattern.

The first
 evidence of azimuthal anisotropy at SPS energies was observed
in the distribution of photons from S+Au collisions at 
200 $\cdot A$ GeV measured in the preshower
photon multiplicity detector of
the WA93 experiment at CERN~\cite{wa93}. 
Since almost 90\% of photons produced in  ultra-relativistic nuclear 
collisions originate from the decay of $\pi^0$'s, the
anisotropy of the observed photon distributions should reflect 
the anisotropy of the $\pi^0$ production followed by the
effects of decay of the $\pi^0$'s. 
Methods have been proposed 
to estimate the anisotropy of the neutral pion emission 
by measuring the anisotropy of photons~\cite{rashmi}.
The decay introduces non-flow correlations between 
the photon pairs due to four-momentum conservation 
and may dilute the correlations between the $\pi^0$'s and the event plane.
Determination of the effect of decay enables the deduction of the
anisotropy of the neutral pions. The photon anisotropy 
measurement thus complements the study of the 
anisotropy of charged particle distributions.

In the present work we report results from the WA98 experiment 
on the centrality dependence of the anisotropy coefficients
extracted from measurement of the azimuthal distributions 
with respect to the event plane of photons and 
charged particles in the same pseudorapidity interval.
Preliminary results on the anisotropy of photon
emission in Pb+Pb collisions have been reported earlier ~\cite{gobinda,sr_icpa}.
The paper is organized in the following manner: Section 2 describes 
the experimental setup and data selection. The analysis technique
and the results on the centrality dependence of the azimuthal anisotropy
of photons and charged particles are discussed in Section 3. 
Section 4 summarises our investigations.

\section{WA98 Experiment and Data Selection}

The WA98 experiment at CERN~\cite{wa98prop} placed emphasis
on  simultaneous detection of hadrons and photons.
The experimental setup consisted of large acceptance hadron and photon 
spectrometers, detectors for photon and charged particle multiplicity 
measurements, and calorimeters for transverse and forward energy 
measurements. The experiment recorded data with 158$\cdot A$ GeV Pb
beams from the  CERN SPS in 1994, 1995, and 1996. The results
presented here are from a portion of the Pb run in 1996 during which
the magnet (GOLIATH) was turned off. The analysis presented here used 
data recorded with the photon multiplicity detector (PMD)
and  the  silicon pad multiplicity detector (SPMD). The data
from the mid-rapidity  calorimeter (MIRAC) was used to characterize
events on the basis of centrality of the collision.

The circular Silicon Pad Multiplicity Detector (SPMD), used for 
measurement of the charged particle multiplicity, was located 32.8 cm from
the target. It had full azimuthal coverage in the region 
$2.35 \le \eta \le 3.75$ (beam rapidity $y_{beam} = 5.81$).
The detector had four overlapping quadrants, each fabricated
from a single 300~{$\mu m$} thick silicon wafer.
The active area of each quadrant was divided into 1012 pads forming 
46 azimuthal wedges and 22 radial bins with pad size increasing with 
radius to provide a uniform pseudo-rapidity coverage. The
intrinsic efficiency of the detector was better than $99\%$.
During the datataking, $95\%$ of the pads worked properly.
The SPMD was nearly transparent to
high energy photons since only about $0.2\%$ are 
expected to interact in the silicon. 
Multiple hits of charged particles on a single pad 
were treated as a single hit for the present analysis, and are counted 
as N$_{hits}$. The detector is sensitive to all charged 
particles. The energy cutoff appears as a low noise
threshold. 
Details of the characteristics of the SPMD
can be found in Refs. ~\cite{WA98-3,spmd_nim}.

The photon multiplicity was measured using the preshower photon 
multiplicity detector (PMD) located at a distance of 21.5 meters
from the target. The detector consisted of 3 radiation length ($X_0$)
thick lead converter plates placed in front of an array of square 
scintillator pads of four different sizes that varied from 
15 mm$\times$15 mm to 25 mm$\times$25 mm, placed in 28 box modules.
Each box module had a matrix of $38\times$50 pads which were 
read out using one image intensifier + CCD camera. 
Details of the design and characteristics of the PMD may be found 
in Ref. ~\cite{pmd_nim,WA98-9}. 

The clusters of hit pads
with a signal above a hadron rejection threshold were 
identified as photon-like. The present analysis has been performed
with the photon-like clusters, which are referred to as photons for brevity. 
Detailed simulations showed that the
photon counting efficiencies for the central to peripheral cases 
varied from $68\%$ to $73\%$. The purity of the photon sample
in the two cases varied from $65\%$ to $54\%$.
Most of the contaminants of the photon sample are charged 
particles which deposit enough energy to fall above the hadron 
rejection threshold. The hadron rejection threshold is taken as  
three times the energy deposited by a minimum ionizing particle. 
For photons this leads to a low p$_T$ threshold of 30 MeV/c.

The transverse energy was measured with the MIRAC calo\-rimeter ~\cite{awes} located at 24.7 meters downstream from the target. 
The MIRAC was used to measure the total transverse energy 
by measurement of both the transverse 
electromagnetic ($E_{\mathrm T}^{em}$) 
and hadronic ($E_{\mathrm T}^{had}$) energies in the 
pseudorapidity interval $3.5\le\eta\le 5.5$. 
The measured total transverse energy,
$E_{\mathrm T}$, provides a measure of the 
centrality of the reaction. 
Events with large $E_{\mathrm T}$ correspond to very central
reactions with small impact parameter, and vice versa.

The minimum bias $E_{\mathrm T}$ distribution has been divided into different
fractions of the minimum bias cross section corresponding to 
different centrality bins ~\cite{WA98-9}.
The most
central selection corresponds to that 5\% of the minimum bias cross section 
$\sigma_{\mathrm MB}$ with largest measured $E_{\mathrm T}$. A
total of about 0.25~Million events have been analysed. The minimum 
number of events in any centrality selection is 15K and the maximum 
is 70K. Table~\ref{data} shows
the percentage cross section and the corresponding number of participants 
for each centrality bin. 
The results presented here use only the data for the pseudorapidity 
region of common coverage of the PMD and SPMD ($3.25\le \eta\le 3.75$) 
where both detectors have full azimuthal coverage. 
The average measured photon
and charged particle multiplicities for this region
of acceptance are also quoted for each centrality in Table~\ref{data}. 

\begin{table}
\begin{center}
\caption{Centrality selections used in the present analysis based on the 
measured total transverse energy. The corresponding fraction of the
minimum bias cross section, number of participants, and the average 
photon and charged particle multiplicities measured in the 
pseudo-rapidity interval $3.25\le \eta\le 3.75$ are given for each
centrality selection.}
\vskip 3mm
\label{data}

\begin{tabular} {lccrr} \hline
 $E_T$(GeV)& \% $\sigma_{MB}$ & N$_{part}$&$<N_{photon}>$&$<N_{hits}>$  \\ \hline 
 40.0- 89.9  &50-80 &  43.7 & 41.1~~ &    34.3~~ \\
 89.9-124.3  &40-50   &87.5   &65.6~~ & 56.9~~  \\
124.3-170.2  &30-40   &123.0   &88.0~~  &78.1~~   \\
170.2-225.5  &20-30 & 172.2 & 116.9~~  &  105.5~~  \\
225.5-298.6  &10-20   & 237.7 &163.1~~   &150.5~~ \\
298.6-347.6  & ~5-10 & 300.4 & 190.1~~ &   177.1~~  \\
$>$347.6      &~0-5~   & 353.4  &222.9~~    &210.3~~ \\

\end{tabular}
\end{center}
\end{table}

\section{Analysis}

The anisotropy of the azimuthal distribution of particle emission
with respect to the reaction plane (or event plane) is characterized 
by the coefficients of the Fourier expansion of the 
azimuthal distribution~\cite{posk}. The first and the second
coefficients are measures of the directed and elliptic flow when the 
expansion is made about the reaction plane, the plane defined by the 
beam direction and the impact parameter.
This may be written as
  
\begin{eqnarray}
\frac{2\pi ~dN}{d(\phi-\psi_R)} = 
1 + 2v_1 \cos(\phi -\psi_R) +  2v_2 \cos 2(\phi-\psi_R) 
\label{fourier}
\end{eqnarray}

\noindent
where $\phi$ is the azimuthal angle of the measured particle 
and $\psi_R$ denotes the azimuthal orientation of the reaction plane.
The reaction plane can be most accurately determined in an experiment 
that measures the (transverse) momenta of the target or 
projectile fragments.
Both $v_1$ and $v_2$ can take positive or negative values.
By convention, positive (negative) values of $v_1$ in 
Eq.~\ref{fourier} denote
flow (anti-flow) in the direction of the deflected projectile fragment, 
and positive (negative) values of $v_2$ indicate
in-plane (out-of-plane) flow. 

Though the most accurate determination of the reaction plane requires the 
measurement of target (or projectile) fragments, most experiments
{\em assume} that the measurement of 
any particle type, in any kinematic window enables a determination of 
the reaction plane. We wish to distinguish between the plane determined 
by projectile or target fragments and the plane determined by any other
particle type, and throughout this article refer to the latter as the 
{\em event plane}. Obtaining the values of coefficients after projecting 
azimuthal angles on the event plane determined from the 
same set (after removing auto 
correlations) maximises the values of anisotropy coefficients, and may 
include non-flow correlations. The values obtained are necessarily 
positive, and have been shown as such in the present work.
The coefficients determined by projecting on an event plane 
from any other set of particles are expected to be 
smaller, and can also have negative values. 
The difference in the values determined using different sets of 
event planes have been included in systematic error by various 
experiments ~\cite{na49,star,phobos}.

\subsection{Method}

In the present analysis,
the azimuthal distributions of particles for any particle species in any
pseudo\-rapidity window is expanded as a Fourier series where the coefficients
of expansion determine the shape of the event. Retaining terms up to 
second order coefficient in the expansion, the shape can be characterized 
by an ellipse for small values of the coefficients. 
The direction of the centroid and
the major axis of the ellipse are determined from the azimuthal
distributions of the particles. These directions, along with the beam 
direction, define the first order and the second order event plane 
respectively, and are obtained as ~\cite{posk}

\begin{eqnarray}
      \psi'_m= \frac{1}{m}\left(\tan^{-1} \frac{\Sigma w_i \sin m\phi_i}
        {\Sigma w_i \cos m\phi_i}\right)
\label{event}
\end{eqnarray}

\noindent where $m$ = 1 or 2 for the first and the second order, respectively.
The $\phi_i$ are the azimuthal angles of the emitted 
particles with respect to a fixed laboratory direction and the 
$w_i$ are the weight factors. For the azimuthal distribution of the 
particle yield, as in Eq.~\ref{fourier}, 
the weight factors are set equal to one. In reality, due
to finite particle multiplicities
$\psi'_1$  and $\psi'_2$ fluctuate about the  
{\em actual} event planes that represent the direction
of the centroid and the direction of the
major axis of the elliptic shape. 
To the extent that initial state nuclear densities are spherically
symmetric and the density fluctuations are negligible, the 
initial nucleon density in the overlap region is 
symmetric about the impact parameter or reaction plane and so
it is expected that the two event plane angles are either the
same or perpendicular to the reaction plane.

The anisotropy, or Fourier coefficients of order $n$,
can be determined from the 
azimuthal distribution of the particles with respect to the event 
plane angle of order $m$, provided $n$ is an integral multiple 
of $m$, by fitting to the following equation~\cite{posk}

\begin{eqnarray}
\frac {dN}{d(\phi-\psi'_m)}& \propto  1 + \sum_{n=1}^{\infty} 2v'_{nm} \cos nm (\phi -\psi'_m)  
\label{firstord}
\end{eqnarray}

\noindent 
$v'_{nm}$ is a measure of the offset of the centroid of 
the distribution when $n \cdot m$ = 1 and is a measure of the difference 
between the major and the minor axes of the ellipse when $n \cdot m$ = 2.
The actual coefficients
are obtained from the observed coefficients $v'_{nm}$ as described later.

Since the event planes do not depend 
on the geometrical setup of the experiment, the distribution of 
the event plane angles determined for a large number of events is expected
to be uniformly distributed in laboratory angle. 
Any non-uniformity in the  acceptance of the detectors 
over the full azimuth will be reflected in a non-uniform distribution of 
event plane angles. 
Any non-uniformity that is identifiable within a large subset of events
can be corrected for by appropriate correction methods. The 
method employed  in the present work is summarized in the following.

\subsection{Detector Acceptance Correction}

The corrected event plane angle is obtained by shifting the 
observed event plane angles $\psi'_m$ by $\Delta \psi'_m$  ~\cite{posk}
where the latter is written as

\begin{eqnarray}
\nonumber {\Delta}\psi'_m = \sum_{n=1}^N\frac{2}{nm}(-&\langle \sin(nm~\psi'_m)\rangle \cos(nm~\psi'_m) + \\
 &\langle \cos(nm~\psi'_m)\rangle \sin(nm~\psi'_m)) 
\end{eqnarray}

\noindent where $N=4/m$ is sufficient to flatten the raw $\psi$ distribution. 
The angular brackets denote an average over all events
and are obtained from the 
raw distribution of the $m^{th}$ order event plane distribution.

\begin{figure*}[bt]
   \centerline{\includegraphics{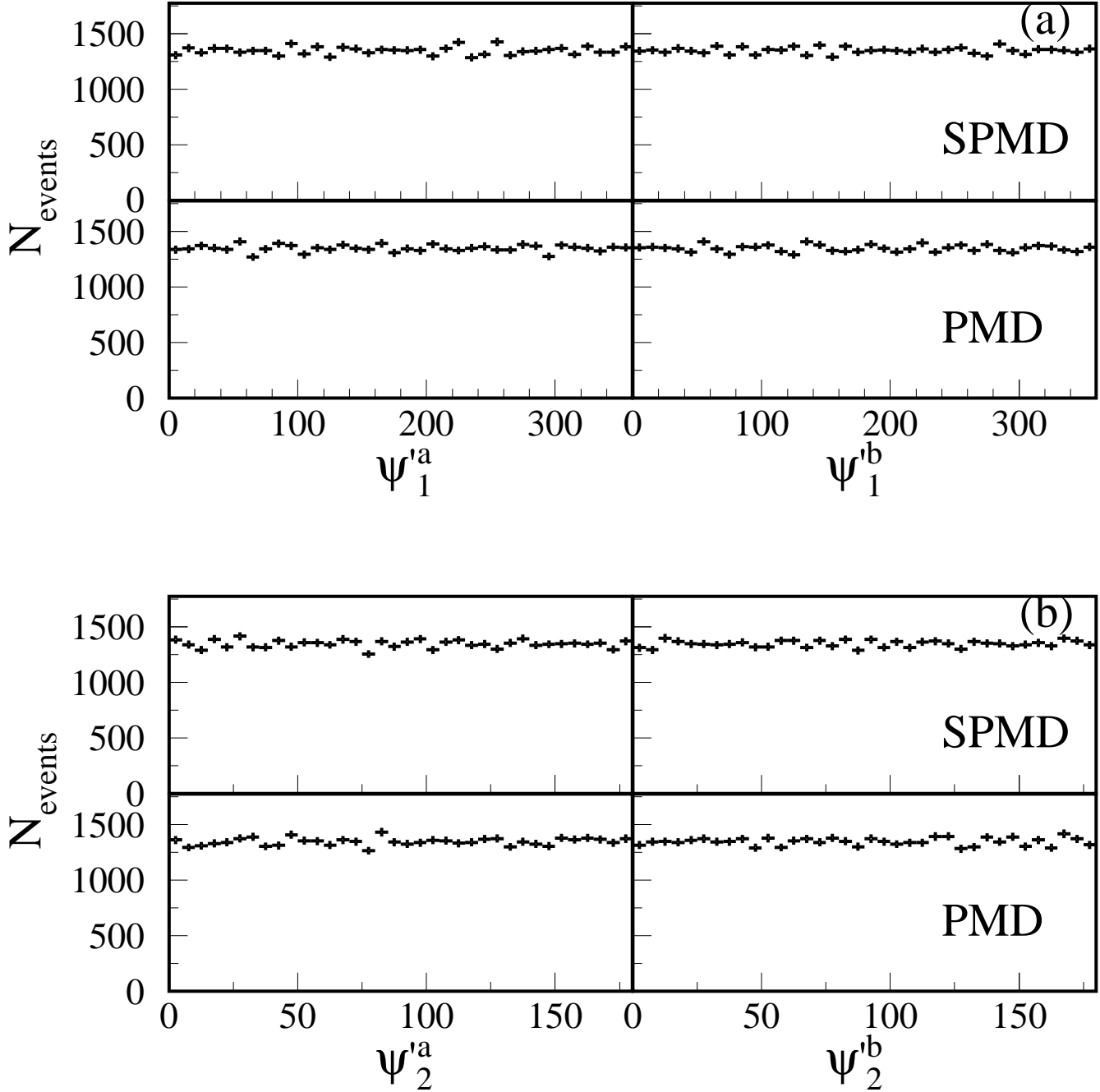}}
   \caption{Acceptance corrected distributions of 
(a) the first order event plane angle, 
$\psi'_1$, of the two subevents for the charged particles 
hits in the SPMD and the 
photons in the PMD in the region 3.25 $\leq \eta \leq$ 3.75 
for the centrality class defined by 225.5 $\leq$ E$_T$ $\leq$ 298.6. 
(b)The same for second order event plane 
angle, $\psi'_2$.
}
   \protect\label{psifig}
\end{figure*}

The distribution of the first and the second order event plane 
angles, corrected for acceptance, is shown in Fig.~\ref{psifig} 
for the charged particle hits in the SPMD and the photon hits 
in the PMD. 

\subsection{Event Plane Resolution Correction}

The average deviation of the estimated event plane 
from the true event plane due to multiplicity fluctuations can be  
determined experimentally and is termed as the resolution correction 
factor (RCF). 
Experimentally, RCF is obtained using 
the  subevent method  described in ~\cite{posk}. 
The particles in each event are sorted in ascending order 
of pseudorapidity. Each event is divided into two equal multiplicity
subevents separated by a pseudorapidity interval.
The two subevents are separated by
one pad for the SPMD and by $\Delta \eta$ = 0.05 for the PMD.
The actual location of the pad in SPMD and the $\Delta \eta$ 
interval in PMD is allowed to vary in the region 3.25 to 3.75 
to ensure equal multiplicity for the two subevents.  
The event plane angle $\psi'_m$ 
is determined for each subevent.\footnote{The distribution of the corrected 
event plane angles for these subevents is observed to be flat, for 
both orders, for photons and for charged particles.}

This enables determination 
of a parameter $\chi_m$ directly from the experimental data
using the fraction of events where the correlation of the planes 
of the subevents is greater than $\pi/2$ ~\cite{posk,ollieconf}:
\begin{eqnarray}
        \frac{N_{events} ( m | {\psi_m'^a - \psi_m'^b|} > \pi/2)}{N_{total}} 
= \frac{e^{-\frac{\chi_m^2}{4}}}{2}
\label{chim}
\end{eqnarray}
where $N_{total}$  denotes the total number of
events, $\psi_m'^a$, $\psi_m'^b$ are the observed  
event plane angles of the two subevents (labeled
$a$ and $b$) and the numerator on the left denotes the
number of events having the angle between subevents greater
than $\pi/2m$. 
The parameter $\chi_m$ so obtained is then used to determine 
RCF$_{nm}$ = $\langle \cos(nm(\psi'_m-\psi^{true}_m))\rangle $,
where $\psi^{true}_m$ is the true direction of the 
event plane, and the average is over all events. 
The RCF can be determined from  $\chi_m$ by the following 
relation from reference~\cite{posk}.

\begin{eqnarray}
 \nonumber \langle \cos(nm(\psi'_m-\psi_m^{true}))\rangle = \frac{\sqrt{\pi}}{2\sqrt{2}} {\chi^{}_{m}} \exp(-\chi^2_m/4) \cdot \\
\left[{{I }}_{\frac{n-1}{2}}(\chi^2_m/4) + {{I}}_{\frac{n+1}{2}} ( \chi^2_m/4 ) \right]
\label{rcfeqn}
\end{eqnarray}

\noindent where I$_\nu$ are the modified Bessel functions of order $\nu$.

The errors on the RCF values have been obtained by considering that
the error on $N_{events}$ is 
statistical. The new 
values of $\chi_m$ are then calculated for values of 
$N_{events} \pm \sqrt{N_{events}}$ 
and used in Eq.~\ref{rcfeqn} 
to calculate new values of RCF. The change in the RCF values
gives the statistical error on the RCF determination. In general, the errors
determined in this way are asymmetric. Symmetric errors have
been quoted using the larger of the two asymmetric errors.

\subsection{Anisotropy Coefficients}

The anisotropy coefficients are obtained by filling up particle
azimuthal distributions of one subevent 
with respect to the event plane of the other subevent where the subevent 
division is described in the previous section.
The anisotropy coefficients have been determined by three methods  
which differ in detail and provide a consistency check.
In this analysis the Fourier
   coefficients prior to event plane
   resolution correction $v'_{nm}$ are extracted for the case with event plane
   order equal to the
   order of the extracted Fourier coefficient, i.e. $v'_{nm}$ = $v'_{nn}$ 
which we will denote by $v'_n$.

In the first method we determine 
$v'^a_n$ = $\langle \cos n (\phi^a - \psi'^b_n) \rangle $  and
$v'^b_n$ = $\langle \cos n (\phi^b - \psi'^a_n) \rangle $ where 
$\phi^a$ represent the azimuthal angles of particles in subevent $a$ and 
$\psi'^b_n$ is the event plane angle determined using particles in
subevent $b$. The averages
are computed over all particles over all events. In the absence of 
non-flow correlations,
$v_n$ can be determined using 
\begin{eqnarray} 
v_n = \sqrt \frac {v'^a_n \cdot v'^b_n}{\langle \cos n (\psi'^a_n - \psi'^b_n) \rangle}
\end{eqnarray} 
This determines
the magnitude of the coefficients and is necessarily positive. 

\begin{figure*}[bt]
   \centerline{\includegraphics{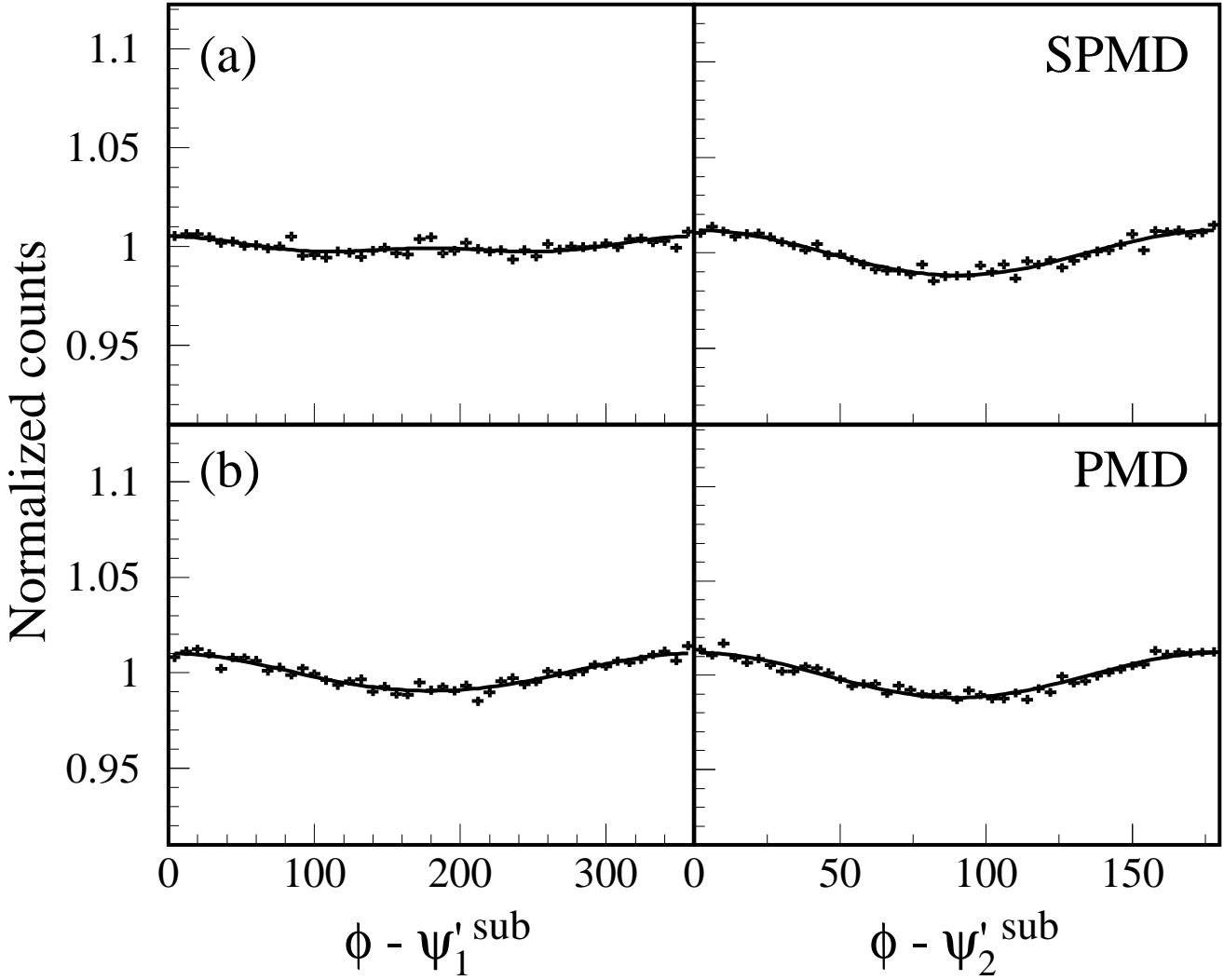}}
\caption{ Distributions of azimuthal angles 
with respect to the first and second order event plane for 
the centrality class defined by 225.5 $\leq$ E$_T$ $\leq$ 298.6.  
(a) for  charged particle hits in the SPMD and (b) for photons in the PMD,
both in the 
pseudorapidity region 3.25 $\leq \eta \leq$ 3.75. 
}
   \protect\label{phiwrtpsi}
\end{figure*}

The distributions with respect to the event plane are shown 
in Fig.~\ref{phiwrtpsi} for both photon and charged 
particles, for both orders for the centrality selection 
corresponding to 10-20\% of the cross section. 
In the second method the 
distributions have been fitted to 
Eq.~\ref{firstord} to determine $v'_{11}$.
The corresponding distributions with respect to the second order event plane
are shown in Fig.~\ref{phiwrtpsi} and have been fitted to Eq.~\ref{firstord} to determine $v'_{22}$. For both orders, the fits have been 
made keeping terms up to values of n $\cdot$ m = 2 in the summation in 
Eq.~\ref{firstord}.

The $v'_{nm}$ are the values determined with respect to the estimated 
event plane and must be corrected for the event plane resolution~\cite{posk}
to obtain the actual anisotropy values 

\begin{eqnarray}
        v_{nm} = \frac{\sqrt{2} \cdot v'_{nm}}{RCF_{nm}} 
\label{corrfact}
\end{eqnarray}
\noindent The factor $\sqrt{2}$ arises because the particle distributions 
have been obtained with respect to the event plane of a subevent with half of 
the total
event multiplicity. The subevent resolution, 
$\sqrt {\langle \cos(nm(\psi^a_n-\psi^b_n))\rangle} $,
is averaged over all events and is in good agreement with RCF$_{nm}/\sqrt{2}$.

In the third method, the values $v_{nn}$ have been obtained directly 
by the subevent method from $\chi_m$ of  
Eq.~\ref{chim} and the fluctuation in the 
average multiplicity M of the full events in 
that centrality bin.

\begin{eqnarray}
        v_{nn} = \frac{\chi_n}{\sqrt{2M}} 
\label{vollie}
\end{eqnarray}

The different methods yield consistent results for both the first order and 
the second order anisotropy coefficients and the difference in the values
is included in the systematic error. The equivalence of the first two 
methods arises due to the equivalence of the geometric mean (method 1) 
and the arithmetic mean (method 2) of the resolution uncorrected values 
$v'_n$ of the two subevents and due to the $1/\sqrt{M}$ dependence of the 
fluctuation in event plane determination. The latter contributes to 
the equivalence of the third method with the first two. 
The three methods may yield different values in the presence of non-flow 
correlations where the differences will be governed by the nature and
strength of these correlations. 
The most probable values are 
determined using method 1 above.
The following sections discuss the systematic effects that distort
the measured anisotropies and the centrality dependence of the 
anisotropies for charged particles and for photons. 

\section{Anisotropy in Charged Particles}

The measured anisotropy is expected to be less than the actual
anisotropy due to the finite granularity of the detector. The 
measured values of anisotropy may also be affected by the 
imprecision in the vertex position due to the finite spread
of the beam. These effects are particularly relevant for the 
charged particle distribution measurement due to the relatively
coarse segmentation and close proximity to the target of the 
SPMD detector. The effects are estimated using simulations. 

\subsection{Granularity}

The finite granularity of the SPMD detector causes a dilution of the 
anisotropy of the azimuthal distribution primarily 
due to efficiency losses from multiple hits.
The quantitative effect of the finite granularity on the measured
anisotropy has been
estimated and is briefly discussed in the following. This has been corroborated
by simulations and is described in greater 
detail in~\cite{grancorr}.

Defining the mean occupancy as the ratio of the number of particles
incident on the detector to the number of active cells  
\begin{eqnarray}
	\mu = \frac{N_{part}}{N_{cell}}
\end{eqnarray}
one can show that
\begin{eqnarray} 
	\mu = ln ( 1 + \frac{N_{occ}}{N_{unocc}})
\end{eqnarray}
where $N_{occ}$ and $N_{unocc}$ are the number of occupied and unoccupied 
cells ~\cite{phoboslow}. One can further show that
\begin{eqnarray}
\frac{N_{hit}}{N_{part}}  
	= \frac{1 - e^{-\mu}}{\mu} 
\label{azim}
\end{eqnarray}
where 	N$_{hit}$ is the number of ocupied cells. 

Since the intrinsic occupancy of cells increases with the increase in 
the number of incident particles, the occupancy will have the same azimuthal
dependence as the incident particles. Substituting 
this for the occupancy in Eq.~\ref{azim} gives us the azimuthal 
dependence of the hits. Expressing the resulting equation as a Fourier
series and collecting coefficients of $\cos n\phi$ enables a determination
of the ratio $v_n^{hits}/v_n$, which to first order can be approximated by
\begin{eqnarray}
 \frac{v_n^{hits}}{v_n} = - \frac{1-x}{x} \cdot {ln(1-x)}
\label{granrat}
\end{eqnarray}
\noindent where $x$ =  $\frac{N_{hit}}{N_{cell}}$ (=$\frac{N_{occ}}{N_{cell}}$).

The anisotropy in the distribution of charged particles can be obtained 
using the measured value of anisotropy in the distribution of hits.

\begin{figure*}[bt]
   \centerline{\includegraphics{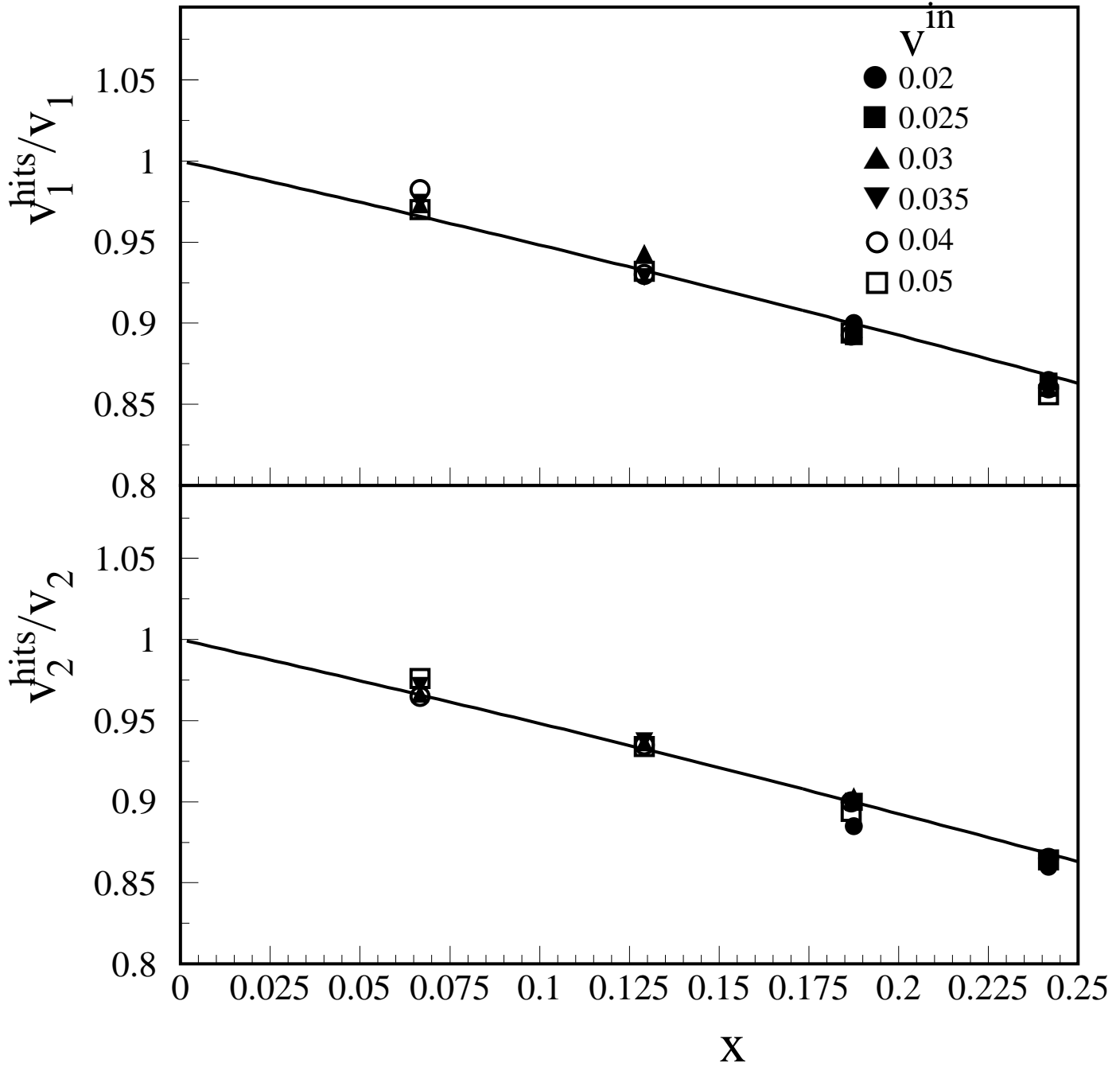}}
\caption{ The ratio $v_n^{hits}/v_n$ for different number of occupied cells.
$x$ is the number of occupied cells scaled with the total number of cells.
Different symbols correspond to different initial values of $v_n$. The solid
line represents Eq.~\ref{granrat}.
}

\label{calib}
\end{figure*}

The azimuthal distributions of the charged particles were 
generated with different initial anisotropies with multiplicities
corresponding to the measured results. The charged particle hits
were sorted into the SPMD bins (2$^\circ$ $\phi$-bin)
assuming a 94\% detection efficiency but taking into account hit losses
due to multiple hits in a single SPMD pad. 
The resulting azimuthal distributions were then analyzed to determine the 
anisotropy coefficients using the method detailed in Sec 3.4 above. 
Fig.~\ref{calib} shows the results of the simulation for both orders, 
where the ratio of the estimated anisotropy from the hit distribution
to the initial anisotropy is shown for varying hit multiplicities, far 
beyond the range of measured values in the SPMD.
This is done for different values of initial anisotropy. 
The correction factor as given by Eq.~\ref{granrat} is shown as 
a solid line in the figure. 
One observes that the simulation results corroborate the results 
obtained above.
The simulation results show
that the extracted anisotropy is 
systematically lower than the initial anisotropy. Part of this loss 
occurs directly due to efficiency losses from multiple hits, and 
contributes both to the anisotropy and the event plane resolution.

\subsection{Shift in Vertex and Beam Spread}

The finite beam size caused an imprecision in the
assumed vertex by up to one mm in the WA98 experiment. 
A small shift in the vertex position does not affect the azimuthal 
distribution in the fine granularity PMD, situated at 21.5 meters 
from the vertex. However, it can
produce an apparent anisotropy in the hit distribution in the SPMD situated 
at 0.328 meters from the target.

The effect of vertex shifts due to the beam spread was also investigated
by simulation.
The vertex position was generated according to a two dimensional
Gaussian with width $\sigma$. Particles were simulated to
originate from this vertex position with a realistic $\eta$ and $p_T$
distribution and an azimuthally symmetric $\phi$ distribution.
The particles were projected onto the SPMD plane, and their hit positions
recorded according to the granularity and nominal location
of the SPMD detector. The recorded positions corresponded to values of 
$\eta$ and $\phi$ which differed from the
generated values due to the shifted position of 
the vertex. The simulated distributions were then analyzed
in the same manner as the experimental data. 
This was repeated for different values of $\sigma$ and 
the values of $v_n$ were obtained for each sample generated. 
The maximum possible shift has been deduced by assuming an azimuthally
symmetric distribution for the most central class, and assigning the 
granularity corrected observed value for first order anisotropy to the 
shift in vertex.
This corresponded to a maximum width of the Gaussian distribution
due to beam spread of 0.3 mm. 

A shift in the vertex position due to beam spread, or time variation 
during the 2.5~s SPS spill, 
produces an azimuthal distribution which has a shifted centroid.
If the shift occurs on an event-by-event basis it 
cannot be corrected for by the acceptance correction methods
discussed above, since they can only correct for average effects.
To investigate the possibility of systematic shifts  
correlated with time during the SPS spill, the SPMD 
charged particle azimuthal distributions were analyzed for 
different times during the spill. No significant variations with
time during the spill  were observed. 

The second order anisotropy is obtained from the 
fit to the elliptical shape of the measured hit distribution.
A vertex shift does not affect the elliptical shape, as verified
by simulations.

\subsection{Results}

\begin{figure*}[bt]
   \centerline{\includegraphics{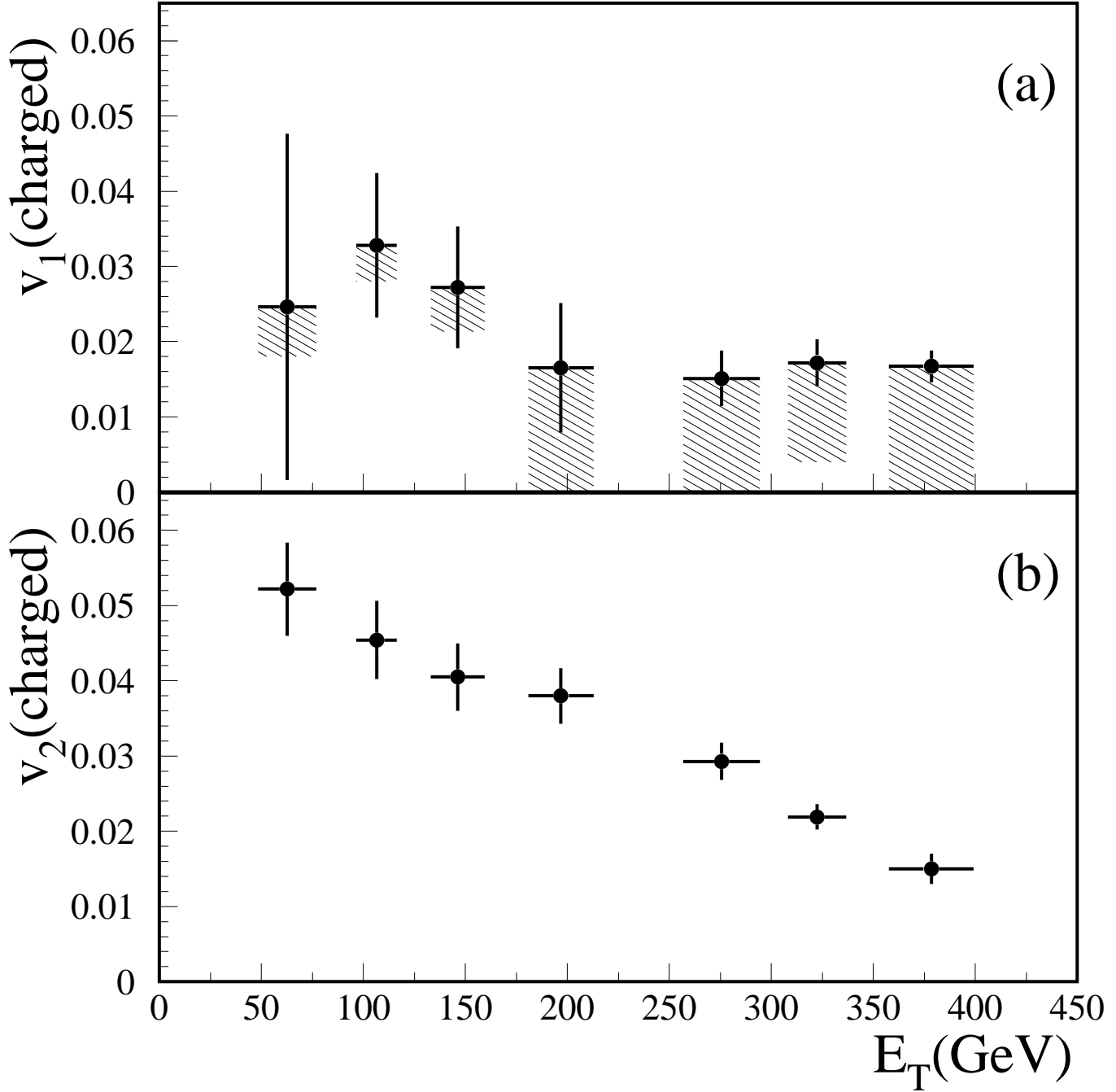}}
\caption{ Anisotropy coefficients of the azimuthal distributions of 
charged particles in the pseudorapidity region 3.25 $\leq \eta \leq$ 3.75 
for different centralities characterised by the measured transverse energy.
(a) First order, $v_1$, where  
the shaded region indicates the extent of the total systematic error 
due to uncertainty in the vertex position.  
(b) second order, $v_2$. 
}

\label{vcharged}
\end{figure*}

The finite granularity 
requires a correction which is obtained using the measured 
values of $v_n$ from the hit distributions and using Eq.~\ref{granrat}.
The corrected results are shown for 
both orders of anisotropy in Fig.~\ref{vcharged}. 
The error bars shown are statistical. 
The results for the first order anisotropy include contributions
from a possible vertex shift. The upper and the lower limit 
of the boxes shown in Fig ~\ref{vcharged}a 
show the asymmetric systematic error and correspond to 
no uncertainty and a maximum uncertainty in the position 
of the vertex, as discussed above.

\begin{figure*}[bt]
   \centerline{\includegraphics{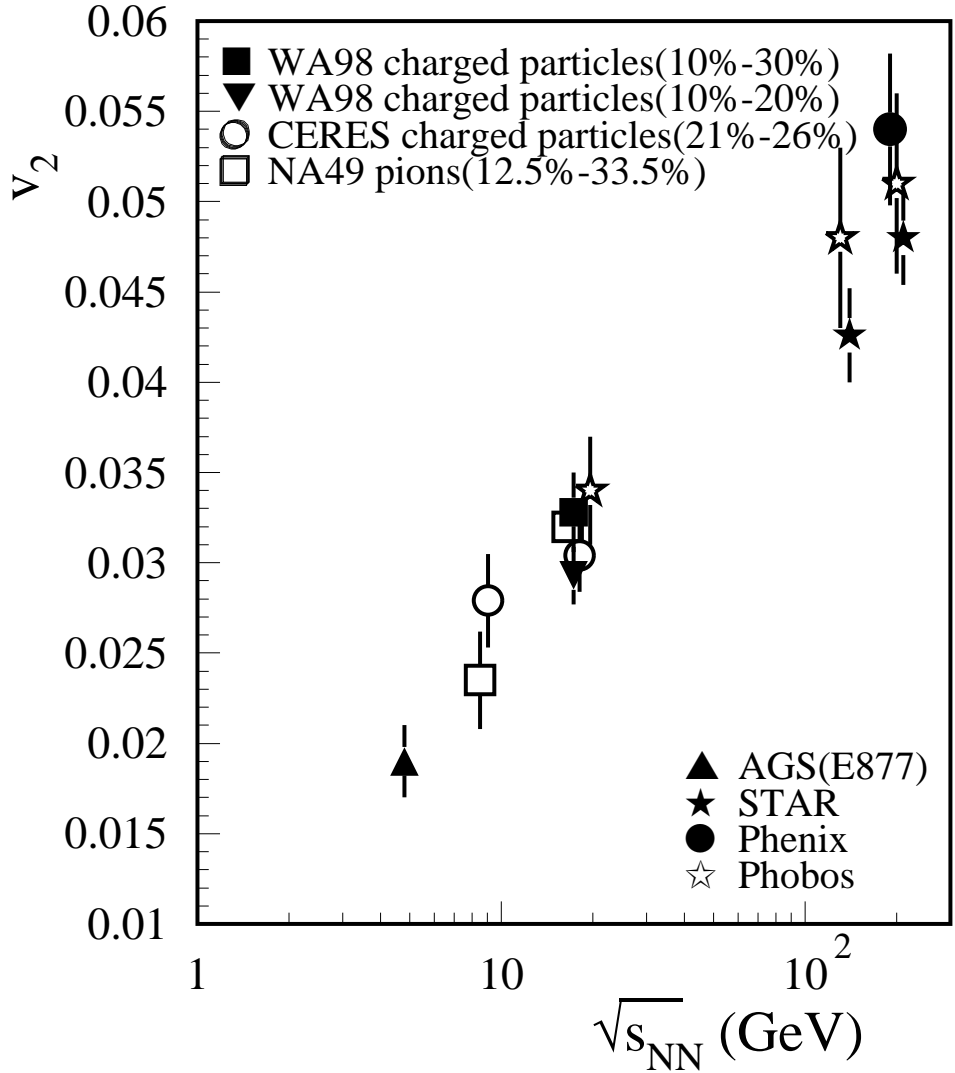}}
\caption{ Second order coefficient for different centre of mass energies.
The results of the present work are for charged particles and are
shown for two different centrality classes. The ordinates of CERES, NA49, STAR
and PHENIX results have been shifted slightly for clarity. The error bars
on WA98 points indicate the statistical and systematical errors added in
quadrature. E877, STAR, PHENIX and PHOBOS $v_2$ values are for 
charged particles. The centrality cuts for all experiments are comparable. 
}

\label{roots}
\end{figure*}

Fig.~\ref{roots} shows $v_2$ as a function of centre of mass energy
from various experiments (results mostly taken from compilation of 
Ref.~\cite{na49pt}). 
These have been measured in different experiments in different kinematic
ranges using methods that vary in detail. The general behaviour shows
a continuous increase in $v_2$ as a function of centre of mass energy.
The results of the present work are shown for two different
centrality ranges corresponding to 10-20\% of cross section and 10-30\%
of cross section. For comparison to the present results, values 
of $v_2$ from PHOBOS, NA49 and CERES experiments 
~\cite{phoboslow,na49pt,ceres} at 
nearly the same centre of 
mass energy are included. The various measurements are seen to agree well 
within errors. 

\section{Anisotropy in Photons}

The photons incident on the PMD predominantly result from the 
two photon decay of the 
neutral pion, and if both photons are detected in the PMD
an additional apparent anisotropy will result from the kinematic
correlation between the photons. 
The limited efficiency and purity of the detected photon sample
in the PMD affect  the measured photon anisotropy values.
The quantitative effect is estimated using simulations and is 
described in the following.

\subsection{Decay Effect}

The decay of neutral pions into two photons introduces correlations 
that can cause apparent anisotropies in the photon distributions 
which are greater than the actual anisotropy of the  
pions. On the other hand, the process of decay smears the photon 
momenta relative to the initial pion momenta and can thereby
dilute the initial correlation 
present in the neutral pions. The relative importance 
of these two competing 
effects has been shown to scale with the experimentally measured quantity
$\chi_m$ and enables a determination of the neutral pion anisotropy
from the measured photon anisotropy~\cite{rashmi}. 
However, the limited efficiency and the contamination of charged particles in 
the sample preclude a determination of the  
$\pi^0$ anisotropy from the measured
$\gamma$ anisotropy in the present work. Using the anisotropy values 
of the charged particles folded with the $\pi^0$ decay and experimental
response allows one to determine the 
expected values of anisotropy of the photons in the PMD, as discussed below.

\subsection{Efficiency and Contamination}

The PMD records 
particle hits which include incident
photons and a contamination of charged particle hits. 
These charged particles
could be primary, or secondary rescattered particles. 
As noted above, the photon counting efficiency (e) of the sample varies 
from 68\% to 73\%  for central and peripheral events 
and the corresponding purity (p) of the sample varies 
from 65\% to 54\%.

The effect of decay, identification efficiency, 
and contamination on the observed photon anisotropy
has been estimated using simulations. The simulations assume that
the photon sample contains a contribution from charged
particle contamination which directly reflects the measured charged
particle anisotropy in addition to a contribution from photons from
$\pi^0$ decays with a $\pi^0$ anisotropy which is also equal to the 
measured charged particle anisotropy.
These simulations use the anisotropy values of the charged particles, 
$\pi^0$ decay kinematics, the PMD acceptance, and the
purity of the PMD photon sample to generate simulated data.
The simulated sample is analyzed to obtain an estimate of the 
expected photon anisotropy corresponding to the observed charged
particle anisotropy.

The neutral pions were generated using the 
experimental pseudorapidity distribution of the charged 
pions ~\cite{wa98klaus} with an exponential p$_T$ distribution
(dN/dp$_T$ = p$_T$ $\cdot$ exp(-6p$_T$)).
The $\pi^0$ multiplicity values were chosen as half of those
measured for charged particles with the SPMD for the same centrality
selection. The second order anisotropy values of $\pi^0$ were
chosen to be linearly increasing with p$_T$ before saturating at
p$_T \sim$ 1.5 GeV/c. For each centrality, the p$_T$ dependence 
was chosen 
to reproduce the p$_T$ integrated mean values of $v_n$ of
charged particles shown in Fig.~\ref{vcharged}.
The linear dependence and the value of p$_T$ at saturation were 
both varied to estimate the systematic errors.
Neutral pions were generated and decayed and the decay photons were
accepted if within the PMD acceptance. 
Using the measured photon multiplicity for a given centrality 
class,  
$p$ $\cdot$ N$_{photons}$ photons were 
randomly selected from those falling onto
the PMD, where $p$ was assigned a value of 0.65 to 0.54 corresponding
to the centrality selection being simulated.  
A background contribution of
$(1-p)$ $\cdot$ N$_{photons}$ charged particles was added to the 
simulated event. 
This simulated data was then 
analyzed using the methods detailed in section 3.4.

The systematic errors in the simulated results have been estimated for 
both orders of anisotropy for each centrality. 
The contribution  to the systematic error on the simulated results are 
shown for one particular centrality selection
(20\%-30\% of minimum bias cross section) in Table~\ref{syserr2}.

\begin{table}[h]
\begin{center}
\caption{Contribution to systematic error from various sources
for the simulated values of both orders of anisotropy of the 
N$_{photons}$ distributions
for the centrality selection corresponding to 20-30\% of minimum 
bias cross section.}
\vskip 3mm
\label{syserr2}
\begin{tabular} {lcc} \hline
Source & First Order & Second order  \\ \hline  
& +0.005 & +0.002 \\
Charged Particle Anisotropy &  &  \\
& $-$0.004 & $-$0.003 \\
Purity of photon Sample &  $\pm$0.002       &   $\pm$0.001     \\
Anisotropy of Contaminants  & $-$0.004     & $-$0.007  \\
$\eta$ and p$_T$ distribution of $\pi^0$  & $\pm$ 0.001  & $\pm$ 0.001 \\
Neutral Pion Multiplicity  &  $\pm$ 0.002     &  $\pm$ 0.001  \\
\\ \hline

\end{tabular}
\end{center}
\end{table}

The relation between the neutral pion anisotropy and photon anisotropy
is not linear ~\cite{rashmi}. $v_n^{PMD}$ has been estimated for a range of 
pion anisotropy
values corresponding to the error measured in this experiment.
An increase (decrease) in the charged particle anisotropy increases 
(decreases) the anisotropy 
in the simulated results of $v_n^{PMD}$ for all centralities.
The percentage change in $v_n^{PMD}$ corresponding to a 10\% change in 
the purity of photon sample is small and is about the same for all 
centralities.
If 50\% of 
the contaminants are assumed to be isotropic, then the resulting anisotropy
decreases for all centralities, and contributes maximumally to the total systematic
error. The uncertainty in the neutral pion multiplicity also has a small 
effect, which is almost independent of centrality.

The systematic error on the measured values of $v_n^{PMD}$ have been 
obtained by 
\begin{itemize}
\item increasing and decreasing the region of acceptance for the analysis
\item varying the size of the interval $\Delta \eta$ between the two 
subevents in the range 0.03 to 0.07. 
\item randomly removing up to 20\% of the photons in the PMD.
\item obtaining the $v_n^{PMD}$ values by the correlation between the 
subevents as described in section 3.4 above.
\end{itemize}
Repeating the analysis by rejecting clusters closer
than twice the size of the scintillator pads did not change the 
$v_n^{PMD}$ values.
 
\subsection{Results}

\begin{figure*}[bt]
   \centerline{\includegraphics{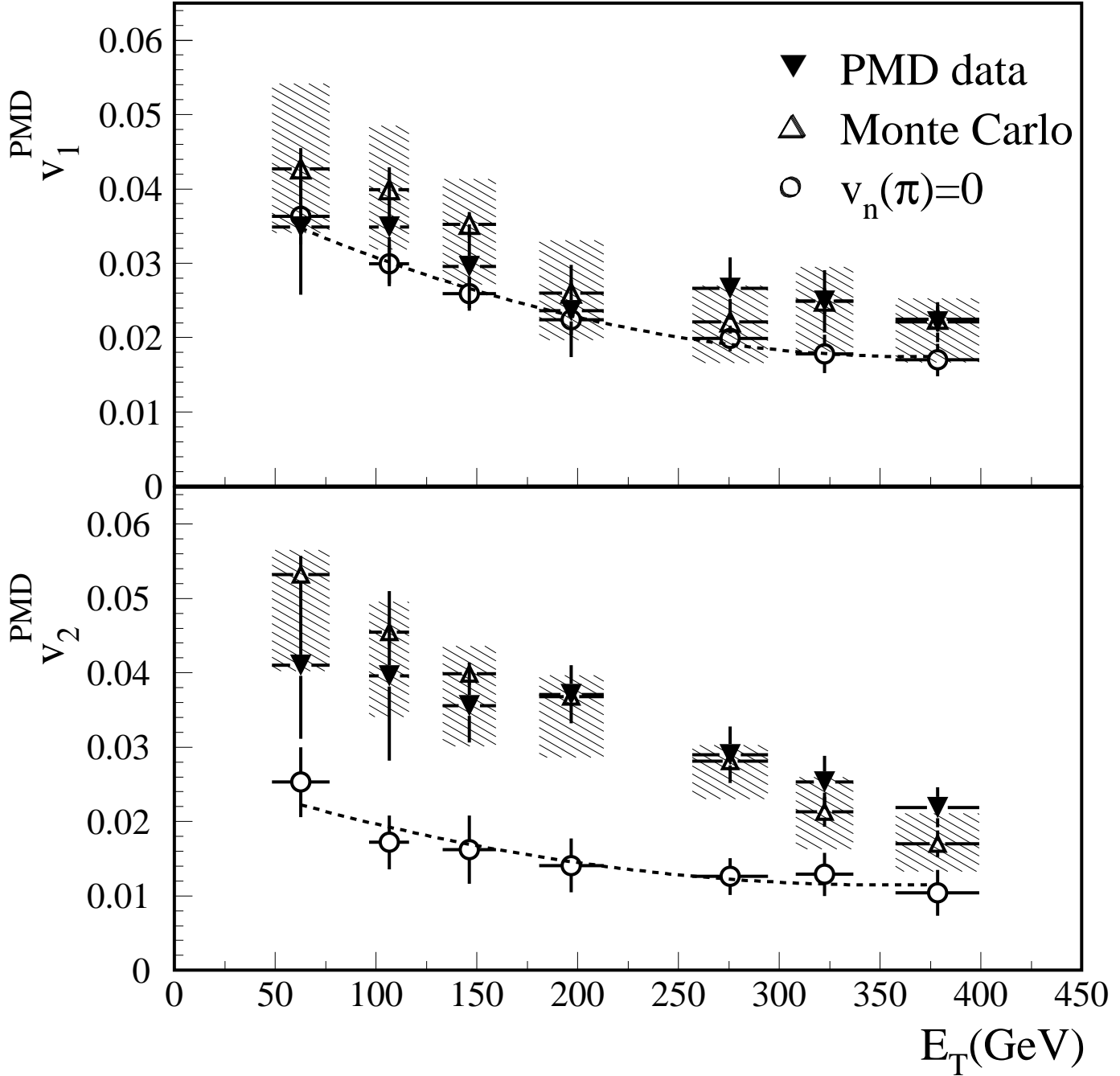}}
\caption{  
 (a) First order, $v_1^{PMD}$, and 
(b) second order, $v_2^{PMD}$, photon anisotropy coefficients 
in the pseudorapidity region
3.25 $\leq \eta \leq$ 3.75 for different centralities are shown by filled
triangles. Statistical and systematical errors are added in quadrature and 
shown as bars on the filled triangles. Open triangles are the 
most probable values of $v_n^{PMD}$ as expected from the simulation.
The shaded regions indicate the simulation uncertainties as described in
section 5.4. The open circles show the calculated values of 
$v_n^{PMD}(v_n(\pi)=0)$
assuming an isotropic distribution of pions with the dashed curve 
indicating a smooth
polynomial fit to the open points. Note, however, that 
$v_n^{PMD}(v_n(\pi)=0)$ can 
not be directly subtracted from the $v_n^{PMD}(v_n(\pi) > 0)$ to obtain the 
anisotropy flow coefficents $v_n$,
as explained in the text.
}

\label{vgamma}
\end{figure*}

 The measured values of $v_n^{PMD}$ are shown in Fig.~\ref{vgamma}. 
The errors on $v_n^{PMD}$ are obtained by adding the systematic and the 
statistical error in quadrature and are also shown.
The open
 triangles show the expected values of $v_n^{PMD}$ from the simulations
 described above. 
The statistical and systematic errors are added in 
quadrature and are shown by the shaded regions.
The 
photon
anisotropy coefficients extracted from the simulated
PMD data are consistent within errors with the 
measured PMD result. 
This demonstrates that the photon anisotropy results measured with the PMD are 
consistent with the charged particle results presented in Sec. 4. Note, 
however, that the results shown in Fig.6 include 
$\pi^0$ decay and charged particle contamination effects, and 
should not be compared directly with other results. In particular, the 
use of the PMD photons themselves
to determine the event plane gives a strong $\pi^0$ decay effect. 
This is shown by
the open circles in Fig.6 which show the simulation values of $v_n^{PMD}$ 
for the assumption of isotropic $\pi^0$ emission, $v_n(\pi)$=0. 
It is seen that the first order photon anisotropy results
are dominated by the $\pi^0$ decay effect while the second order 
photon anisotropy results 
also have a significant
decay contribution, which cannot simply be subtracted from the measured $v_2^{PMD}$. 
Since the relationship between $v_{\gamma}$ and $v_{\pi^0}$ 
depends in a non-trivial way
on the $\pi^0$ ($p_T$ dependent) multiplicity and 
anisotropy ~\cite{rashmi,wa98leda} it is non-trivial
to extract the $\pi^0$ anisotropy from the photon anisotropy measured with 
respect to the event plane determined using photons.
It should be noted that 
the WA98 photon $v_n$ results measured with the LEDA photon detector 
covering the photon rapidity region $2.3 \le ~y \le 2.9$
~\cite{wa98leda} were obtained with 
respect to an event plane determined in the target fragmentation region. 
The $v_n^{PMD}$ measured using the PMD photons and $v_n$ measured 
using the LEDA photons will therefore have different sensitivity 
to the $\pi^0$ decay effect. The PMD results include effect of decay 
correlations, and therefore the current measurement, $v_n^{PMD}$, 
provides upper limits on the 
anisotropic flow coefficients $v_n$.

\section{Summary}

The azimuthal angle distributions
with respect to the event plane have been measured for charged particles
and photons with full azimuthal coverage
in the pseudorapidity region of 3.25 $\le~ \eta~ \le$ 3.75 for
 $^{208}$Pb + $^{208}$Pb collisions at 158 $\cdot A$ GeV/$c$.
A total of 0.25 million
events, classified in seven centrality selections, have been used in the 
analysis. The Fourier coefficients of the azimuthal 
distributions have been extracted in several ways, all
giving consistent results
for the first order $v_1$ (directed) and 
$v_2$ second order (elliptic) anisotropies
for photons and charged particles.
The results show the expected trend of decreasing anisotropy with increasing
centrality for both $v_1$ and $v_2$  for charged particles and photons. 
Our results 
agree with the 
results reported by other experiments for near similar conditions. 
~\cite{phoboslow,na49pt,ceres,wa98leda}.
The observed anisotropies of the photon distributions  
compare well with those obtained from simulations
that include the charged particle contamination and the
correlations arising due to the decay of the neutral pions 
assumed to have the same anisotropies as measured 
for charged particles.  

\begin{acknowledgement}

\begin{sloppypar}
We wish to express our gratitude to the CERN accelerator division for the
excellent performance of the SPS accelerator complex. We acknowledge with
appreciation the effort of all engineers, technicians, and support staff who
have participated in the construction of this experiment.

This work was supported jointly by
the German BMBF and DFG,
the U.S. DOE,
the Swedish NFR and FRN,
the Dutch Stichting FOM,
the Polish KBN under Contract No. 621/E-78/SPUB-M/CERN/P-03/DZ211/,
the Grant Agency of the Czech Republic under contract No. 202/95/0217,
the Department of Atomic Energy,
the Department of Science and Technology,
the Council of Scientific and Industrial Research and
the University Grants
Commission of the Government of India,
the Indo-FRG Exchange Program,
the PPE division of CERN,
the Swiss National Fund,
the INTAS under Contract INTAS-97-0158,
ORISE,
Grant-in-Aid for Scientific Research
(Specially Promoted Research \& International Scientific Research)
of the Ministry of Education, Science and Culture,
the University of Tsukuba Special Research Projects, and
the JSPS Research Fellowships for Young Scientists.
ORNL is managed by UT-Battelle, LLC, for the U.S. Department of Energy
under contract DE-AC05-00OR22725.
The MIT group has been supported by the US Dept. of Energy under the
cooperative agreement DE-FC02-94ER40818.
\end{sloppypar}
\end{acknowledgement}

\pagebreak

\pagebreak

\pagebreak

\pagebreak

\pagebreak

\pagebreak

\end{document}